\documentclass[journal]{IEEEtran}
\usepackage{graphicx}
\graphicspath{{figures/}}

\usepackage{caption}
\usepackage{subcaption}
\usepackage[table,xcdraw]{xcolor}
\usepackage{multirow}
\usepackage{amsmath,environ}
\usepackage{dblfloatfix}

\ifCLASSINFOpdf
\else
\fi
\begin{document}
%
\title{A Cost \& Performance-Efficient Field-Programmable Pin-Constrained \\Digital Microfluidic Biochip}
%
%
%

\author{Alireza~Abdoli and~Ali~Jahanian
\thanks{A. Abdoli is with the Department
of Computer Science and Engineering, Shahid Beheshti University, G.C., Tehran, 1983963113 Iran e-mail: al.abdoli@mail.sbu.ac.ir}
\thanks{A. Jahanian is with the Department
of Computer Science and Engineering, Shahid Beheshti University, G.C., Tehran, 1983963113 Iran e-mail: jahanian@sbu.ac.ir}
}
\maketitle

\begin{abstract}
Digital microfluidic biochips (DMFBs) constitute modern generation of Lab-on-Chip (LoC) devices aimed at automation, miniaturization and cost-affordability of biochemistry and laboratory procedures. Over the course of past few years there have been various application-specific and general-purpose DMFBs aimed at reduced manufacturing costs; following the same trend this study presents a general-purpose DMFB with highly competitive characteristics compared with the state-of-the-art DMFBs. The proposed DMFB architecture provides lower Layout/PCB fabrication costs thereby reducing the total manufacturing costs. While more cost-affordable the proposed design is competitive with the state-of-the-art DMFB architectures.
\end{abstract}


%
\IEEEpeerreviewmaketitle

\section{Introduction}
%
%
%
%

\IEEEPARstart{M}{icrofluidic} biochips are modern revolutionary devices enabling a new paradigm in performing fluidic bio-chemistry and laboratory manipulations never existed before; providing various ranges of applications among which are DNA multiplexed Polymerase Chain Reaction (PCR) \cite{1}, in-vitro diagnostics \cite{2} and protein crystallization \cite{3} and DNA computing \cite{58}.

The conventional bio-chemistry operations are mostly performed/controlled by human intervention. On the other hand traditional methods of accomplishing laboratory operations require considerable amounts of experiment materials and reagents; which might be costly with experimental testing of new drugs. Yet, laboratory equipment consume large amounts of space thus a room might be dedicated to accommodation of the aforementioned equipment. Also, automated robotic laboratory equipment cost much more than affordable to an end-user person.

In order to address aforesaid issues the microfluidic biochips were developed; these chips are mainly aimed at three crucial factors of automation, cost-affordability and miniaturization.

Microfluidic biochips are accompanied by a microcontroller used for programming the chip to perform wide ranges of laboratory procedures along with the bio-chemistry operations without any human interventions thus realizing automation factor. These chips operate on the basis of manipulating negligible amount of fluids thus providing significant cost-affordability in terms of experiment materials and reagent consumption. Yet, these chips are manufactured at scales resulting in much smaller area consumption as compared to the conventional circuits \cite{59} and laboratory equipment; this implies the miniaturization factor inherent in the microfluidic biochips.

\begin{figure}[t]
\centering
\subcaptionbox{}[.3\linewidth][c]{%
\includegraphics[scale=0.15]{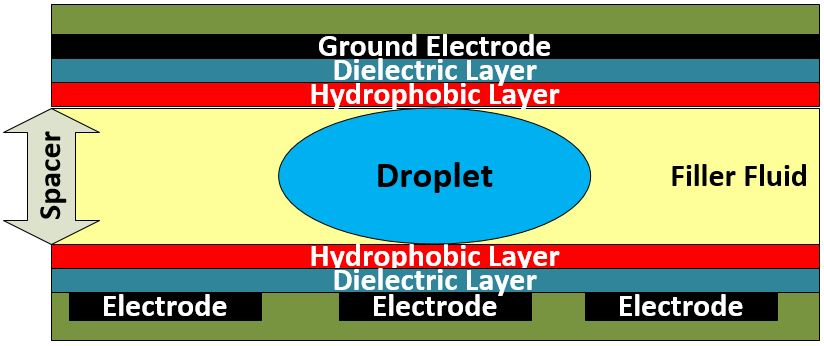}}\quad \quad \quad \quad
\subcaptionbox{}[.3\linewidth][c]{%
\includegraphics[scale=0.2]{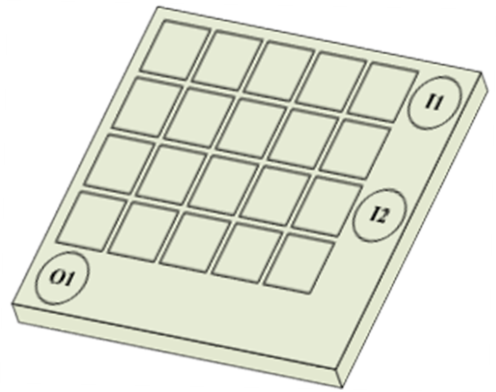}}
\caption{(a) Cross-sectional and (b) top view of a closed DMFB}
\label{f2}
\end{figure}

A typical digital microfluidic biochip consists of two plates; the bottom plate is consisted of an array of equal-size electrodes while the top plate spans the bottom plate and acts as the ground electrode. Figure \ref{f2} illustrates the top and cross-sectional view of a DMFB.

The electrodes forming the array of electrodes are connected to the pins of the microcontroller; the microcontroller can be programmed in order to turn on/off the electrodes such that the intended bio-assay is realized. Droplets are sandwiched in between the top and bottom plates; at the bottom plate there are electrodes on which droplets would be actuated. Considering the structure of the bottom plate on top of electrodes there is a dielectric layer; additionally a hydrophobic layer is placed on top of the dielectric layer. The hydrophobic layer is used for facilitating movement of droplets on the surface of the DMFB. In order to ease movement of droplets further the space between the top and bottom plates is filled with some filler fluid, typically silicon oil, which allows for better movement of droplets on the array of electrodes.

This paper presents a low-cost yet general-purpose digital microfluidic biochip; the proposed design is considerably smaller than the previous state-of-art designs which in turn results in significant improvements in terms of total number of electrodes and control pins used for driving electrodes, along with cheaper fabrication costs; also the proposed improvements in the hardware design results in shorter droplet routing times thus improving the overall bio-assay completion times.

The rest of the paper is organized as follows: section II is devoted to review of underlying technologies and the design flow associated with digital microfluidic biochips. Section III provides literature review of the previous works on DMFB designs. Section IV initially reviews the original general-purpose field-programmable DMFB design on the basis of which the enhanced design proposed in this study is established. Section V is devoted to presentation of simulation results in comparison with other previous notable designs. Eventually, section VI concludes the paper.

\section{DMFB TECHNOLOGY AND DESIGN FLOW}

This section initially reviews the fundamental technologies related with digital microfluidic biochips, then proceeding to the DMFB design flow.

\subsection{DMFB Technology Overview}

\begin{figure}[b]
\centering
\subcaptionbox{}[.3\linewidth][c]{%
\includegraphics[width=1in]{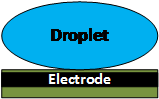}}\quad
\subcaptionbox{}[.3\linewidth][c]{%
\includegraphics[width=1in]{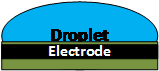}}
\caption{Electrowetting on Dielectric phenomenon}
\label{f1}
\end{figure}

Digital microfluidic biochips are based on the electro-wetting on dielectric (EWOD) phenomenon \cite{4}; which is the electromechanical actuation (wetting) of conductive fluids on a solid surface through electrical bias \cite{5}. As a result of which droplets are actuated by applying appropriate level of voltage to the desired electrode; thus creating an electrical field affecting the droplet over the activated electrode. This happens because of the tension between the droplet and the electrode; Figure \ref{f1} depicts the electro-wetting on dielectric phenomenon.

As depicted in Figure \ref{f1} in case no voltage is applied the droplet remains in its normal form whereas in case appropriate level of voltage is applied the droplet is polarized; this phenomenon can be put to work for moving the droplets on the array of electrodes. It must be noted that currently droplets can be actuated to neighboring (top, bottom, left and right) electrodes; though droplets currently cannot be actuated to diagonally adjacent electrodes.

\begin{figure*}[!t]
\centering
\subcaptionbox{Storage}[.15\linewidth][c]{%
\includegraphics[width=0.75in, height=0.75in]{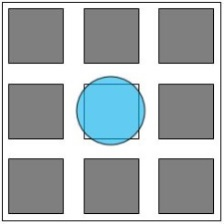}}\quad
\subcaptionbox{Transportation}[.15\linewidth][c]{%
\includegraphics[width=0.75in, height=0.75in]{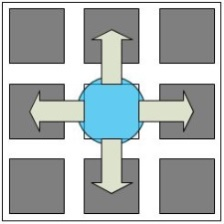}}\quad
\subcaptionbox{Merging}[.15\linewidth][c]{%
\includegraphics[width=0.75in, height=0.75in]{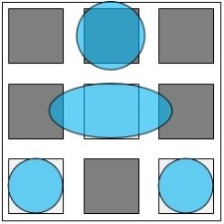}}\quad
\subcaptionbox{Splitting}[.15\linewidth][c]{%
\includegraphics[width=0.75in, height=0.75in]{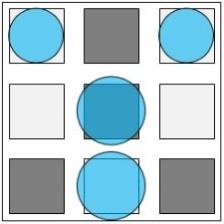}}\quad
\subcaptionbox{Mixing}[.15\linewidth][c]{%
\includegraphics[width=0.75in, height=0.75in]{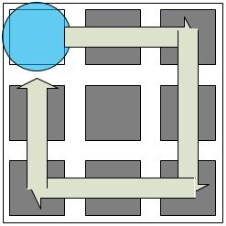}}\quad
\subcaptionbox{Heat/Detect}[.15\linewidth][c]{%
\includegraphics[width=0.75in, height=0.75in]{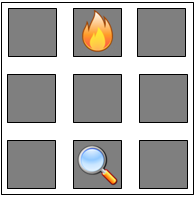}}
\caption{Fundamental microfluidic operations (Please note that light gray electrodes are active whereas dark gray electrodes are deactivated)}
\label{f3}
\end{figure*}

A typical DMFB is capable of various microfluidic operations including dispensing, transporting (movement), merging, mixing, and also detection/heating; Figure \ref{f3} depicts various fundamental microfluidic operations.

The most fundamental microfluidic operation is holding (storing) the droplets which is achieved by activating the electrode beneath the droplet. On a large DMFB at any given time there might be several droplets on the DMFB; which must be held steady for the duration of their presence on the DMFB.

Second fundamental microfluidic operation is transporting (moving) the droplets on the array of electrodes. As stated earlier, droplets can be moved to adjacent (top, bottom, left or right) electrodes. This is achieved by deactivating the electrode beneath the droplet and activating the desired neighboring electrode.

Third fundamental microfluidic operation corresponds to merging two droplets into a single larger droplet. Initially the two droplets are moved near each other; in this case there is a distance of one electrode between the two droplets. Then the electrode between the two droplets is activated while at the same time deactivating the two electrodes holding droplets. This causes both droplets to be moved towards the just activated electrode; thus merging the two droplets into a single larger droplet.

Fourth fundamental microfluidic operation is splitting a single droplet into two smaller, ideally equal sized, droplets. This is accomplished by activating neighboring electrodes (top/bottom or left/right) while at the same time deactivating the electrode beneath the droplet. This splits the droplet into two smaller, ideally equal sized, droplets.

Fifth fundamental microfluidic operation is heating/detection/cooling which requires availability of external equipment in the DMFB. Given the architecture of the DMFB external heaters/detectors/coolers are affixed on top of designated electrodes during the manufacturing process so that enabling aforementioned capabilities.

\subsection{DMFB Synthesis Flow}

Performing any bioassay on the DMFB involves various stages referred to as synthesis flow of the DMFB. Initially, the process is initialized by inputting the protocol of the bioassay and also architecture specifications of the DMFB.

\begin{figure*}[!b]
\centering
\includegraphics[width=\textwidth, height=1.5in]{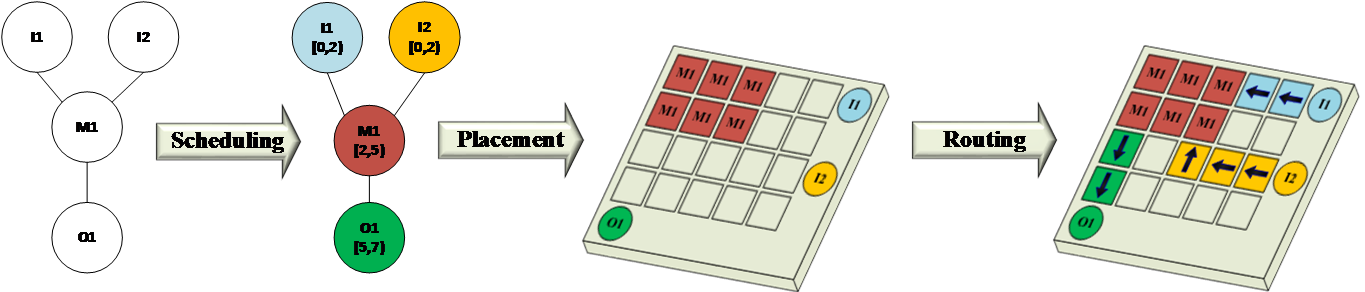}
\caption{Synthesis flow of a typical digital microfluidic biochip (Please note that letter I denotes input while letter O denotes output reservoirs; also letter M represents mixing operations)}
\label{f4}
\end{figure*}

The protocol of a given bioassay is in the form of a directed acyclic graph (DAG). On the other hand, the architecture specification of the DMFB incorporates information on dimensions of the array of electrodes, locations of I/O reservoirs on the periphery of the array of electrodes and also location of any fixed modules (e.g. detectors/heaters/coolers). Figure \ref{f4} illustrates several stages of the DMFB synthesis flow.

\subsubsection{Scheduling}

Given the protocol of the bioassay and also the architecture specifications the synthesis flow is commenced by scheduling microfluidic operations within the bioassay protocol. Scheduling is the first stage of the synthesis flow during which every microfluidic operation is assigned with exact start and end times. The scheduling algorithm must make the best use of available resources to ensure microfluidic operations are scheduled as efficient as possible thus producing the shortest overall scheduling time.

\subsubsection{Placement}

Following the scheduling stage every microfluidic operation has exact start and end times. Next, the scheduled operations must be placed on the array of electrodes according to specific resource type required by the operation. The placement algorithm must be performed such that all scheduled operations during any given time-step are successfully placed on the array of electrodes. There are two main categories of placement algorithms; namely, free placement and fixed placement algorithms.

In free placement the algorithm keeps account of available spaces on the array of electrodes and attempts to find appropriate area to place the operation on the array of electrodes. On the other hand, in case of fixed placement locations of modules on the array of electrodes are already specified; thus there is no need to search for available areas. This reduces the placement to a binding problem; in which operations are bound to first available module. The free placement is more time-consuming computationally; yet, free placement of modules might be performed such that there is no space for movement of droplets in between the placed modules thus causing droplet blockages and deadlocks.

In case of fixed placement given the already fixed location of modules there are dedicated spaces for routing of droplets between the modules thus guaranteeing successful droplet routing and eliminating the possibility of blockages and deadlocks associated with free placement algorithms.

\subsubsection{Droplet Routing}

There are several scenarios involved in the droplet routing stage; dispensing droplet into modules, transportation of droplet between the modules and outputting of droplet from modules to output reservoirs. The droplet routing algorithm must operate such that the shortest possible route is produced. Also, the task must be accomplished such that the droplet does not interfere with other droplets already present on board. Furthermore, droplets must be routed such that not any two droplets collide with each other while being routed to their respective destinations.

\subsubsection{Pin-Mapping}

Early DMFBs applied direct-addressing scheme \cite{35} in which every single electrode is dedicated with an independent control pin; this provides highest degree of flexibility and controllability, yet the scheme requires highest number of control pins. As an illustration a DMFB of size m $\times$ n requires m $\times$ n control pins which might be excessively large in case of large DMFBs.

A cheaper alternative to direct-addressing scheme would be cross-referencing scheme \cite{36}; in which every single column and row is assigned with a controlling pin; thus driving an array of size m $\times$ n requires m + n control pins. Yet, simultaneous activation of multiple rows and columns might cause unintended droplet movement.

\begin{figure}[t]
\centering
\subcaptionbox{}[.3\linewidth][c]{%
\includegraphics[width=0.75in, height=0.75in]{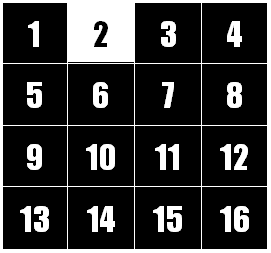}}\quad
\subcaptionbox{}[.3\linewidth][c]{%
\includegraphics[width=0.75in, height=0.75in]{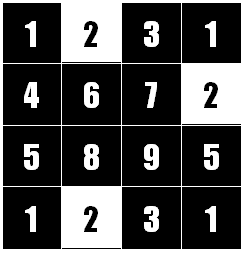}}
\caption{(a) direct-addressing (b) pin-constrained pin-mapping schemes}
\label{f5}
\end{figure}

Figure \ref{f5} shows schematic of direct-addressing and pin-mapping schemes for 4 $\times$ 4 DMFB designs. As can be seen in Figure \ref{f5}, the direct-addressing scheme provides the highest level of flexibility; however requires highest number of pin-count for addressing the array of electrodes.

A more promising option is the active-matrix scheme \cite{37} which is capable of driving m x n electrodes just by m + n control pins; the scheme allow the highest level of flexibility just as with the direct-addressing scheme and further eliminating challenging associated with cross-referencing scheme. While requiring lower number of control pins this approach is not stable enough to be widely used.

Given the high cost of direct-addressing scheme, limitations of cross-referencing scheme and instability of active-matrix scheme researchers proposed the pin-constrained scheme \cite{35} \cite{38}; in which initially electrodes with common functions are grouped together. Next, electrodes of the same group are assigned with a single shared pin thus significantly reducing overall number of control pins. Although, reducing the total number of control pins the approach comes with reduced degree of flexibility and controllability compared with direct-addressing and active-matrix schemes. On the other hand the pin-constrained scheme demands significantly lower number of pin-count; as a result of reduced pin-count the level of flexibility is diminished. Given the advantages and disadvantages of aforesaid pin-mapping schemes the pin-constrained scheme comes up as the most accessible choice for manufacturing of DMFBs.

\subsubsection{Wire Routing}

The wire routing stage deals with the wirings from microcontroller pins to the array of electrodes. Given the type of pin-mapping the amount of wire routing would vary. This directly affects the number of metal-layers used toward the wire routing stage; thus significantly affecting the overall manufacturing cost of the DMFB.

\section{Previous Works}

This section reviews major previous works addressing various recent DMFB architectures and algorithms. Over the course of past few years there have been numerous studies on DMFBs.

Xu et al. \cite{35} proposed a broadcast-addressing DMFB; their proposed design encompasses a multi-function pin-constrained DMFB capable of executing a predefined set of bioassays. Luo et al. \cite{47} proposed a pin-constrained pin-mapping scheme allowing concurrent movement of two droplets on the DMFB; their proposed design is targeted towards performing a predefined set of bioassays.

Keszocze et al. \cite{48} proposed a DMFB synthesis based on their general and exact routing methodology. Their proposed design results in significantly low number of control pins. While demanding small number of control pins their proposed design is solely applicable to small to mid-size bio-assays (e.g. PCR, In-vitro diagnostics); this is because the computational time for large complex bio-assays such a protein-split would be prohibitively large.

Wille et al. \cite{49} proposed a one-pass synthesis scheme with much faster computation times compared with \cite{48}; faster computation times allow for performing large bio-assays such as protein-split not computationally feasible in \cite{48}. Although much more efficient in terms of computation times the runtime of the produced solutions is approximate and considerably variable among different runs.

Abdoli et al. \cite{31} \cite{57} proposed a field-programmable pin-constrained design; their proposed design is considerably smaller than previous pin-constrained designs also yielding reduced overall number of electrodes and controlling pins. This paper enhances the DMFB design in \cite{31}. Also, Abdoli et al. \cite{32} proposed their architecture with a cellular structure providing a regular expandable structure; their proposed structure is inspired by the widely popular FPGA devices.

Grissom et al. \cite{33} proposed a field-programmable pin-constrained design aimed at general-purpose bioassay execution. Also, Grissom et al. \cite{34} proposed their enhanced low-cost pin-constrained DMFB design so that requiring less number of metal-layers towards wire-routing of the DMFB; thus, yielding significantly reduced overall manufacturing costs.

\section{Pin-Constrained Scheme}

The architecture layout and pin-mapping scheme have a great impact on the overall manufacturing costs of the design. The field-programmable pin-constrained designs \cite{31} \cite{32} \cite{33} \cite{34} employ a layout and scheme such that to provide general-purpose bioassay execution along with reduced pin-count so that the overall manufacturing cost of the DMFB is significantly reduced. Given the considerable number of electrodes allocated to droplet routing paths then efficient addressing of routing paths is of significant importance in terms of pin-count and overall DMFB manufacturing costs.

\begin{figure}[!b]
\renewcommand{\arraystretch}{1.3}
\centering
\includegraphics[width=\columnwidth]{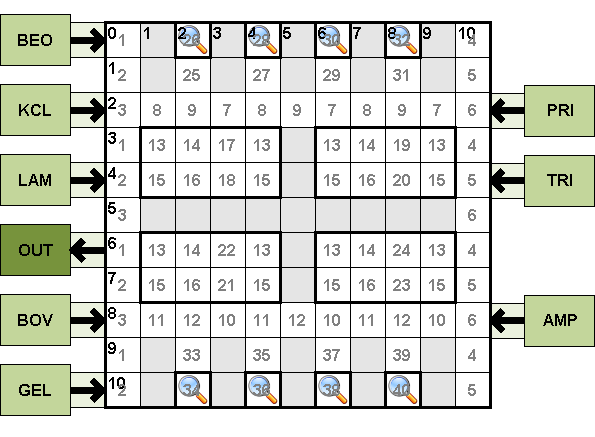}
\begin{tabular}{|l|c|}
\hline
\multicolumn{1}{|c|}{Description} & Pins \\
\hline
\multicolumn{2}{|c|}{\cellcolor[HTML]{000000}{\color[HTML]{FFFFFF} Routing Paths}} \\
\hline
Left Vertical Routing Path & 1 $-$ 3 \\
Right Vertical Routing Path & 4 $-$ 6 \\
Top Horizontal Routing Path & 7 $-$ 9 \\
Bottom Horizontal Routing Path & 10 $-$ 12 \\
\hline
\multicolumn{2}{|c|}{\cellcolor[HTML]{000000}{\color[HTML]{FFFFFF} Modules}} \\
\hline
Mixing module (Shared) & 13 $-$ 16 \\
Mixing module I/O & 17 $-$ 20 \\
Mixing module Hold & 21 $-$ 24 \\
SSD module I/O & 25 $-$ 32 \\
SSD module Hold & 33 $-$ 40 \\
\hline
\end{tabular}
\caption{Simulation of pin-constrained pin-mapping scheme for the proposed DMFB design}
\label{f6}
\end{figure}

A typical general-purpose pin-constrained DMFB design consists of mixing modules, storage/split/detection (SSD) modules and droplet routing paths towards movement of droplets; simulation pin-mapping scheme of the proposed DMFB design is depicted in Figure \ref{f6}.

Looking at Figure \ref{f6} there are 4 mixing modules and 8 SSD modules; also there are two vertical routing paths at the left and right of the design; used towards inputting/outputting droplets into/out of I/O reservoirs. Also, there are two horizontal routing paths at the top and bottom of the design; which allow for inputting/outputting droplets into/out of mixing and SSD modules. Mixing modules are used for merging and mixing of droplets; while SSD modules are used for storing, splitting and detection of droplets.

\subsection{Pin-Constrained Operation Execution}

A typical general-purpose pin-constrained DMFB must be capable of performing various microfluidic operations involving droplet dispensing/outputting, merging, mixing and splitting/heating/detection.

\subsubsection{Droplet Dispensing/Outputting}

The typical bioassay execution on a DMFB initially involves droplets being dispensed from I/O reservoirs on the perimeter of the array of electrodes. Each I/O reservoir has an individual electrode leading a droplet to the edge of the array of electrodes \cite{33}; since common to all DMFB designs these electrodes are omitted from DMFB designs.

\subsubsection{Droplet Merging/Mixing}

Droplet merging operation within pin-constrained DMFB designs is performed inside mixing modules; during which initially the first droplet fluid is moved into an available mixing module and then moved to the Hold pin of the module; then the second droplet is moved to the I/O electrode of the same module containing the first droplet while at the same time moving the first droplet off the Hold electrode towards the I/O electrode. Moving the first and second droplets to the I/O electrode causes merging of the two droplets. After successfully merging the two droplets the resulting droplet is moved back to the Hold electrode of the mixing module; this completes the merging operation. Then the droplet is rotated around the module for a certain period of time so that contents of the droplet are perfectly mixed; this accomplishes the mixing operation.

\subsubsection{Storage/Splitting/Detection}

Sometimes it might happen that droplets produced by some operations need to wait on-chip for certain number of time-steps before being used towards other microfluidic operations. SSD modules can be used for storing droplets; every SSD module is capable of storing one droplet at any time-step.

Furthermore, SSD modules can be used for splitting a droplet into ideally two equal size droplets. This is achieved by moving a droplet onto the I/O electrode of an available SSD module; then the I/O electrode is deactivated while at the same time activating the Hold electrode of the SSD module and also the routing electrode leading to the SSD module. This ideally causes the droplet to be split in half thus producing two smaller droplets. Then one of the droplets is stored in the current SSD module while the second droplet is moved to another available SSD module. Additionally, in case equipped with external detector/heater/cooler the SSD module can be used for performing detection/heating/cooling operations.

\subsection{Pin-Constrained Droplet Routing}

Droplet routing is consisted of a set of tasks in order to move droplets from input reservoirs to mixing/SSD modules, between mixing and SSD modules and from mixing/SSD modules to output reservoirs. Given the considerable number of electrodes devoted to routing paths using an individual pin per electrode requires large pin-count which in turn requires additional hardware for driving routing path electrodes. In order to reduce the pin-count associated with routing paths the 3-phase routing path is used \cite{50}. The 3-phase routing, as the name implies, requires only 3 pins for addressing any routing paths of arbitrary length. It must be noted that the 3-phase pin-mapping can be used in case of intersecting routing paths; however, this requires using different pin numbers for addressing intersecting routing paths in order to avoid conflicts and unintended droplet movements. Looking at Figure \ref{f6} the 3-phase routing is used for pin-assignment of routing paths in the design. As stated earlier, different pins are used for different routing paths to avoid conflicts and unintended movement of droplets.

\section{THE PROPOSED DMFB DESIGN FLOW}

This paper enhances the field-programmable pin-constrained DMFB design originally proposed in \cite{31}. The original design offered numerous advantages compared with previous designs among which are smaller array dimensions, lower number of electrodes and also reduced number of controlling pins. Lower dimensions results in significantly lower droplet routing times such that the total bioassay execution times are considerably reduced.

\subsection{The Proposed DMFB Design}

This section introduces several improvements compared with the original GFPC DMFB design \cite{31} which makes the proposed design competitive to the state-of-art pin-constrained DMFB designs.

\subsubsection{Improved Pin-Mapping For Routing Paths}

The original GFPC design used pin numbers 1-3 for vertical routing paths and pin numbers 4-6 for horizontal routing paths; while efficient in terms of pin-count this complicates the wire-routing of the original GFPC design which results in increased number of metal-layers used towards wire-routing of the design. Considering the significance of pin-mapping on the wire-routing stage of the DMFB synthesis flow the proposed design utilizes a revised pin-mapping scheme towards routing paths; in which every routing path is assigned with 3 dedicated pins. While increasing the pin-count this greatly simplifies the wire-routing of the proposed DMFB design which in turn results in decreased number of metal-layers used towards wire-routing of the design. The revised pin-mapping of the proposed design for routing paths is illustrated in Figure \ref{f6}.

\subsubsection{Improved Pin-Mapping For Mixing Modules}

\begin{figure}[t]
\centering
\subcaptionbox{Original GFPC}[.3\linewidth][c]{%
\includegraphics[scale=0.25]{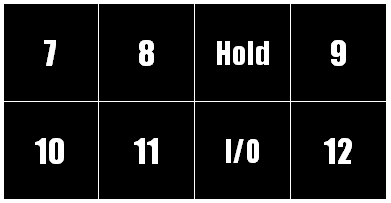}}\quad \quad \quad
\subcaptionbox{Proposed Design}[.3\linewidth][c]{%
\includegraphics[scale=0.25]{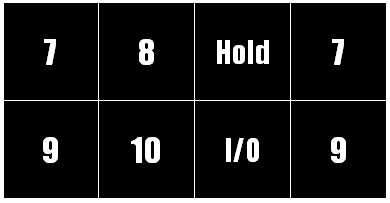}}
\caption{Pin-assignment of mixing modules in the original GFPC design [31] compared with the proposed design}
\label{f7}
\end{figure}

The original GFPC design \cite{31} allocated a single shared pin to every electrode in the mixing modules. The scheme required 6 pins to address the electrodes inside mixing modules; yet the proposed pin-assignment method for mixing modules reduces the number of mixing pins from 6 to only 4 pins. Figure \ref{f7} depicts the pin-mapping of mixing modules in \cite{31} versus the pin-mapping of this study. As illustrated in Figure \ref{f7} (a) 6 pins (7$-$12) are used for mixing pins; whereas the enhanced pin-mapping proposed in this study requires only 4 pins (7$-$10) for addressing electrodes inside mixing modules.

Yet, as a disadvantage it must be mentioned that applying the proposed pin-mapping scheme for mixing modules results in increased power consumption within mixing modules; Rotating a given droplet in a mixing module with the original GFPC design required eight electrode actuations; whereas in case of the proposed design it requires 10 electrode actuations (that is because as illustrated in Figure \ref{f7} (b) activating pins 7 and 9 causes redundant electrode actuations). Thus power consumption of mixing modules in the proposed design is increased by 25\% compared with the original GFPC design \cite{31}.

\begin{figure}[b]
\centering
\subcaptionbox{Overall}[.3\linewidth][c]{%
\includegraphics[scale=0.25]{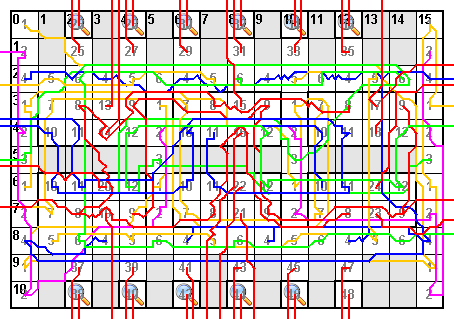}}\quad \quad \quad
\subcaptionbox{Layer 1}[.3\linewidth][c]{%
\includegraphics[scale=0.25]{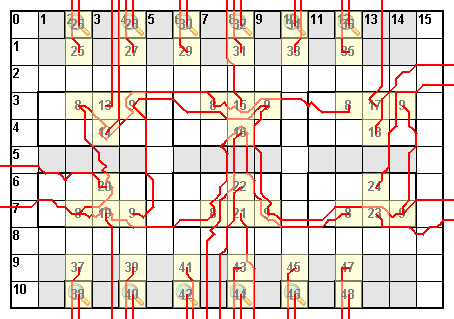}} \\

\subcaptionbox{Layer 2}[.3\linewidth][c]{%
\includegraphics[scale=0.25]{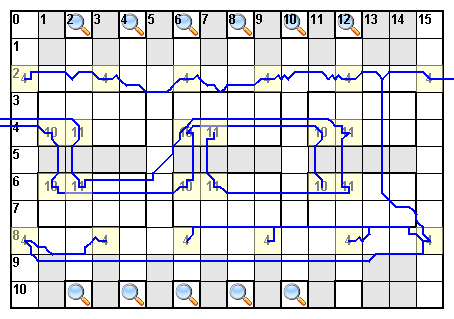}}\quad \quad \quad
\subcaptionbox{Layer 3}[.3\linewidth][c]{%
\includegraphics[scale=0.25]{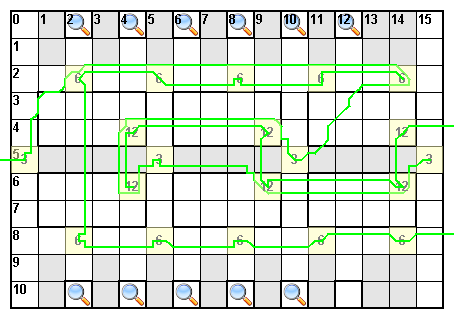}} \\

\subcaptionbox{Layer 4}[.3\linewidth][c]{%
\includegraphics[scale=0.25]{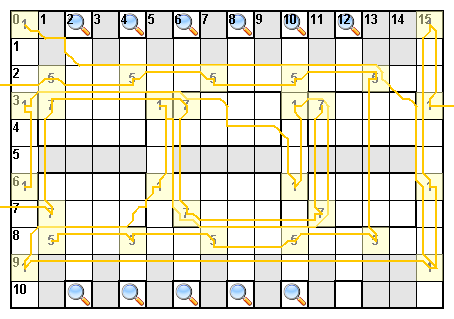}}\quad \quad \quad
\subcaptionbox{Layer 5}[.3\linewidth][c]{%
\includegraphics[scale=0.25]{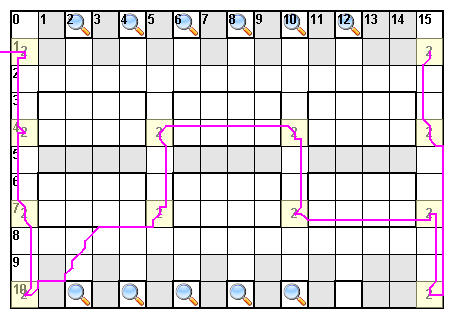}}
\caption{Wire-routing of the original GFPC architecture \cite{31} with orthogonal capacity of 2}
\label{f8}
\end{figure}

\subsubsection{Reduced Overall Manufacturing Costs}

\begin{figure}
\centering
\subcaptionbox{Overall}[.3\linewidth][c]{%
\includegraphics[scale=0.25]{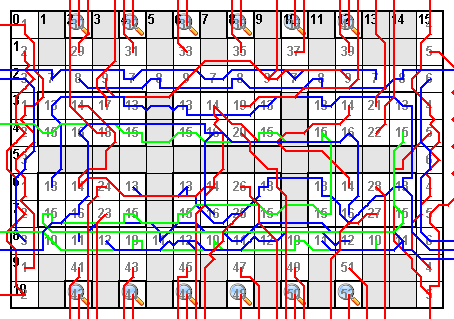}}\quad \quad \quad
\subcaptionbox{Layer 1}[.3\linewidth][c]{%
\includegraphics[scale=0.25]{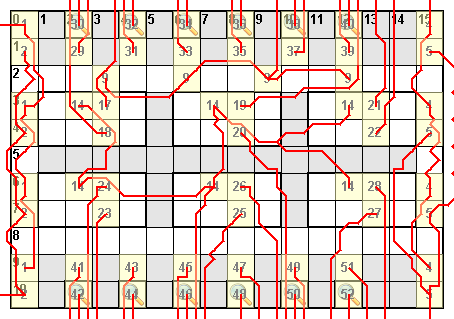}} \\

\subcaptionbox{Layer 2}[.3\linewidth][c]{%
\includegraphics[scale=0.25]{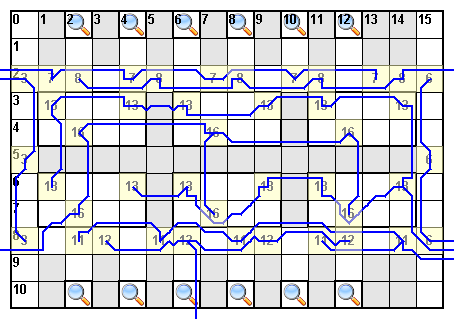}}\quad \quad \quad
\subcaptionbox{Layer 3}[.3\linewidth][c]{%
\includegraphics[scale=0.25]{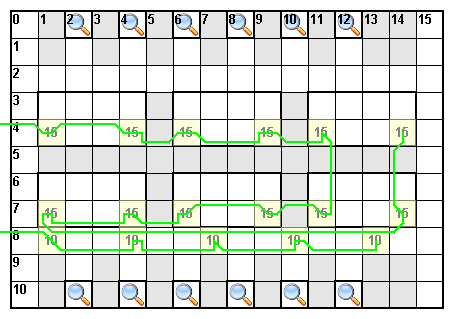}}
\caption{Wire-routing of the proposed design}
\label{f9}
\end{figure}

Given the improved pin-assignment of the proposed design the wire-routing of the design is improved such that the total number of metal-layers required for wire-routing of the proposed design is reduced from 5 to 3 layers; while retaining the functionalities of the original GFPC design. This reduction in the total number of metal-layers directly affects the overall manufacturing cost of the design. Figure \ref{f8} is devoted to the wire-routing of the original GFPC design \cite{31} while Figure \ref{f9} illustrates the improved 3 layers wire-routing of the proposed design in this study.

\subsubsection{Improved Fault-Tolerance}

Given the significant pin-count reduction provided with pin-constrained DMFB designs the level of flexibility is negatively affected. The proposed design allocates different ranges of pin numbers for addressing routing paths throughout the design. Though, allocating different pins for different routing paths increases the total pin-count however it significantly improves the flexibility of the design. Because of the availability of various different routing paths a faulty electrode in a given routing path can be tolerated and bypassed through other routing paths.

Abdoli et al. \cite{51}\cite{56} investigated fault-tolerance of the original GFPC DMFB design. A major advantage of the design compared with previous pin-constrained DMFB designs lies with the availability of various routing paths; which allows bypassing any faulty electrodes hindering the way of droplet towards the intended destination.

\subsubsection{Improved Power Consumption in Routing Paths}

The proposed design enjoys improved power consumption within droplet routing paths. The original GFPC design \cite{31} assigned two distinct 3-phase routing paths (i.e. total pin-count of 6) for addressing all droplet routing paths throughout the design. In case of the proposed design a distinct3-phase pin-mapping is dedicated to every individual routing path (i.e. total pin-count of 12 as illustrated in Figure \ref{f6}); albeit the increased pin-count, assigning distinct 3-phase pin-mapping to individual routing paths results in decreased power consumption within droplet routing paths. The improved pin-mapping applied to droplet routing paths of the proposed design results in at least 50\% in terms of power consumption within droplet routing paths compared with original GFPC \cite{31} design.

\subsubsection{No Need for Routing Buffer (RB) Module}

Prior general-purpose pin-constrained designs in \cite{31} \cite{32} \cite{33} \cite{34} allocated a routing buffer (RB) module for bypassing possible mutual droplet routing deadlocks. Since designs in \cite{33} \cite{34} used a single routing path for droplet routing it is necessary to allocate an RB module to resolve possible routing deadlocks. In case of the proposed DMFB design since there are two groups of top and bottom mixing/SSD modules connected by vertical droplet routing paths at the left and right of the design there is no need to allocate separate resources in the form of routing buffer module. This saves design space and pin-count; eliminating the RB module reduces the total pin-count by 2.

\subsection{Synthesis Flow of the Proposed DMFB Design}

This section briefly discusses various algorithms used towards synthesis flow of the proposed DMFB design.

\subsubsection{Scheduling}

For the scheduling stage of the proposed design the list scheduling algorithm \cite{2} a fast however greedy algorithm is applied.

\subsubsection{Placement}

Given the already fixed location of modules the placement stage of the design flow is reduced to a binding problem during which microfluidic operations are bound to modules with already-fixed locations. The binding phase of the proposed design is performed using Grissom’ left-edge binding algorithm for field-programmable pin-constrained DMFB designs \cite{20}.

\subsubsection{Droplet Routing}

The droplet-routing stage of the proposed design works on the basis of the sequential droplet routing algorithm originally proposed for the GFPC DMFB design \cite{31}; the revised pin-mapping of the proposed design differs with the original GFPC design in that the routing paths connecting the upper groups of mixing/SSD module to the lower group is omitted from the proposed design. Omitting inter-module droplet routing paths significantly simplifies wire-routing stage of the design flow. Though, the minor difference the overall structure still remains the same thus the sequential routing algorithm proposed for the original GFPC design still applies the proposed DMFB design.

\subsubsection{Pin-Mapping}

The pin-mapping of the proposed design is depicted in Figure \ref{f6} for 11 $\times$ 11 architecture. The left vertical routing column is allocated with pins 1-3. Next, the right vertical routing column is assigned with pins 7-9. In case of horizontal routing columns the pins 10-12 are assigned (as can be seen in Figure \ref{f6}), the proposed design accommodates two horizontal routing columns; intended for providing access to mixing/SSD modules located at the top and bottom tiers of the proposed design. Next, pin-mapping of mixing modules is addressed; in which initially shared pins are assigned with pins 13-16. As discussed earlier pin-mapping of the proposed design solely requires 4 pins for addressing shared pins inside mixing modules; then every mixing module must be allocated with two independent pins for addressing I/O and Hold pins of the module. Looking at Figure \ref{f6} it can be seen that pins 17-20 are used for addressing Hold pins of mixing modules while pins 20-23 are allocated towards I/O pins of mixing modules. Given allocation of pins to mixing modules then the process continues with pin-mapping of SSD (Storage/Split/Detection) modules; which merely consists of two pins (I/O and Hold pins). As can be seen in Figure \ref{f6} there are 8 SSD modules with pins 25-32 assigned to I/O pins and pins 33-40 assigned towards Hold pins.

\subsubsection{Wire Routing}

Given the improved pin-mapping of the proposed design compared with the original GFPC design \cite{31} the wire-routing stage of the design flow is simplified so that the total number of metal-layers is reduced to 3 layers. The wire-routing stage of the proposed DMFB design utilizes the negotiated-congestion wire-router algorithm \cite{42}.

\section{HARDWARE COST ANALYSIS}

The overall manufacturing cost of a typical DMFB is affected by various factors among which are:

\begin{itemize}
\item Dimensions of array of electrodes
\item Number of metal-layers
\item Pin-count
\end{itemize}

In this section it is attempted to provide detailed cost information on manufacturing costs of the proposed DMFB design while at the same time comparing with the state-of-art designs already available; this helps to show how the proposed DMFB design retains original capabilities while at the same time remaining competitive with the state-of-art DMFB designs \cite{31} \cite{34}.

\begin{figure}[b]
\centering
\includegraphics[scale=0.2]{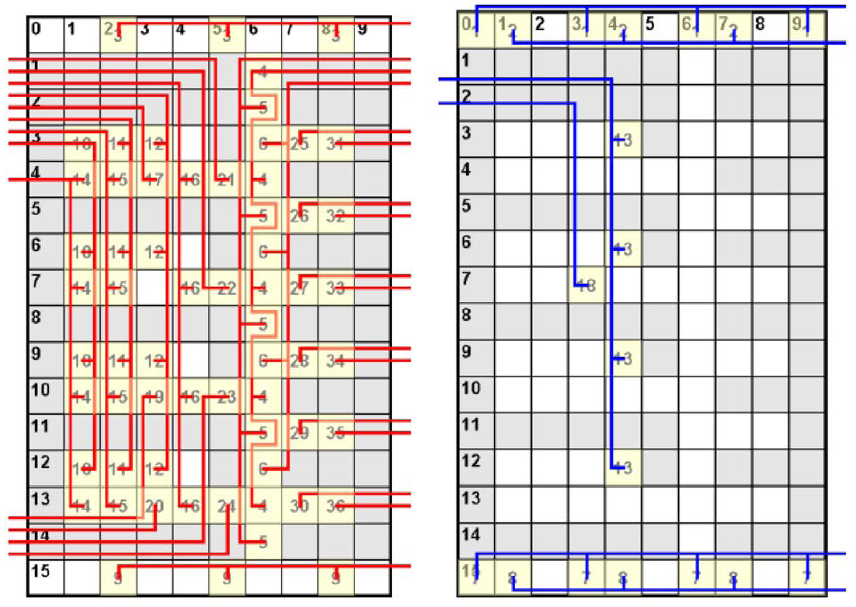}
\caption{Wire-routing of the Enhanced FPPC DMFB design \cite{34} given orthogonal capacity of 2 (First and second layers are denoted with red and blue wires, respectively)}
\label{f11}
\end{figure}

The original FPPC DMFB design \cite{33} proposed a field-programmable pin-constrained DMFB design capable general-purpose bioassay execution. The Enhanced FPPC DMFB \cite{34} proposed by Grissom et al., is a general-purpose DMFB design requiring a one/two metal-layer(s) towards wire-routing; the number of metal-layers in \cite{34} varies given the orthogonal capacity parameter applied during the wire-routing stage. In case of orthogonal capacity equal to 2 the wire-routing stage yields a two metal-layers wire-routing. Figure \ref{f11} illustrates the two metal-layers wire-routing of their proposed DMFB design.

Table \ref{t1} shows a summary on characteristics and wire-routing details of various general-purpose DMFB designs. The column ‘Name’ denotes name of the DMFB design; while column ‘WR Alg.’ refers to the wire-routing algorithm applied to the DMFB design. The column ‘Array Dim.’, along with sub-columns ‘X’ and ‘Y’ show array dimensions of the DMFB design. Columns ‘\# Elec’ and ‘\# Pin’ denote number of electrodes and control pins in each design, respectively. The column ‘\# Metal-Layers per Orthogonal Capacity’ represents number of metal-layers used towards wire-routing given different orthogonal capacities.

Considering results of Table \ref{t1} it is seen that the original FPPC \cite{33} design yields 5 metal-layers solution with orthogonal capacity of 2; on the other hand the enhanced FPPC \cite{34} yield 2 metal-layers solution for the orthogonal capacity of 2. Increasing the orthogonal capacity to 3 achieves 1 metal-layer solution for the enhanced FPPC.

\begin{table}[t]
\renewcommand{\arraystretch}{1.3}
\setlength{\tabcolsep}{3.0pt}
\centering
\caption{Illustrating number of wire-routing metal-layers for various orthogonal capacities}
\label{t1}
\begin{tabular}{|l|c|c|c|c|c|ccccccc|}
\hline
\multicolumn{6}{|c|}{DMFB Details} & \multicolumn{7}{c|}{} \\ \cline{1-6}
\multicolumn{1}{|c|}{} & & \multicolumn{2}{c|}{\begin{tabular}[c]{@{}c@{}}Array\\ Dims.\end{tabular}} & & & \multicolumn{7}{c|}{\multirow{-2}{*}{\begin{tabular}[c]{@{}c@{}}\# Metal-Layers per \\ Orthogonal Capacity\end{tabular}}} \\ \cline{3-4} \cline{7-13} 
\multicolumn{1}{|c|}{\multirow{-2}{*}{Name}} & \multirow{-2}{*}{\begin{tabular}[c]{@{}c@{}}WR\\ Alg.\end{tabular}} & X & Y & \multirow{-2}{*}{\# Elec} & \multirow{-2}{*}{\# Pin} & \multicolumn{1}{|c|}{2} & \multicolumn{1}{c|}{3} & \multicolumn{1}{c|}{4} & \multicolumn{1}{c|}{5} & \multicolumn{1}{c|}{6} & \multicolumn{1}{c|}{7} & \multicolumn{1}{c|}{8} \\ \hline
Original FPPC {\cite{33}} & {\cite{42}} & 12 & 33 & 237 & 63 & \cellcolor[HTML]{000000}{\color[HTML]{FFFFFF} 5} & 5 & 6 & 6 & 5 & 6 & 5 \\
Enhanced FPPC {\cite{34}} & {\cite{34}} & 10 & 30 & 148 & 65 & 2 & \cellcolor[HTML]{333333}{\color[HTML]{FFFFFF} 1} & 1 & 1 & 1 & 1 & 1 \\
Original GFPC {\cite{31}} & {\cite{42}} & 21 & 11 & 169 & 58 & 5 & 5 & 5 & \cellcolor[HTML]{333333}{\color[HTML]{FFFFFF} 4} & 4 & 5 & 4 \\
FPCA {\cite{32}} & {\cite{42}} & 21 & 17 & 235 & 58 & 7 & \cellcolor[HTML]{333333}{\color[HTML]{FFFFFF} 6} & 7 & 6 & 7 & 7 & 6 \\
Proposed Design & {\cite{42}} & 16 & 11 & 122 & 52 & \cellcolor[HTML]{333333}{\color[HTML]{FFFFFF} 3} & 3 & 3 & 3 & 3 & 3 & 3 \\ \hline
\end{tabular}
\end{table}

The remaining of this section discusses detailed cost analysis of the wire-routing stage; for the sake of fairness and similarity solely, the Enhanced FPPC \cite{34} and the proposed design will be addressed.

As stated in earlier the proposed DMFB design requires 3 metal-layers wire-routing; this is obviously more than the two metal-layers wire-routing solution for Enhanced FFPC DMFB design. However, given the larger dimensions of the Enhanced FPPC DMFB design for accommodating equal number of resources (mixing/SSD modules) compared with the proposed DMFB design in this work and also higher pin-count compared with the present work the overall manufacturing cost of the proposed DMFB design would be competitive with the Enhanced FPPC DMFB design.

To our best knowledge authors in \cite{34} for the first time ever attempted to provide detailed information on overall manufacturing costs of DMFB designs. Considering the overall cost of a typical DMFB design the first key element is the number of metal-layers used for wire-routing. The second and third key elements are architecture dimensions and pin-count number, respectively. Larger architecture dimensions mean higher PCB costs; on the other hand higher number of pin-count requires more equipment for driving those extra pins which again costs extra charges to the overall manufacturing costs of the DMFB.

The authors in \cite{34} used Advanced Circuits’ online instant quote feature \cite{52} to obtain estimations on the cost of PCB according to dimensions of the architecture. Also, according to \cite{34} it is assumed that DMFBs are driven by an Atmega 1284 microcontroller with 32 general-purpose I/Os (GPIOs) \cite{53}. In case a DMFB design requires more than 32 pins then additional circuitry is needed which accomplished by daisy-chaining arbitrary number of Fairchild 74VHC595MTC 8-bit shift registers \cite{54}; quantities of 2500 of the aforementioned shift register can be purchase for \$0.14 per unit from Mouser \cite{55}. The following equation obtained from \cite{34} shows the number of shift registers required for driving any DMFB designs.

\begin{equation}
\begin{split}
\#ShiftRegs = \\
& \left\{ \,
\begin{IEEEeqnarraybox}[][c]{l?s}
\IEEEstrut
\left \lceil \frac{numPins - 28}{8} \right \rceil & numPins $>$ 32, \\
0 & otherwise.
\IEEEstrut
\end{IEEEeqnarraybox}
\right.
\end{split}
\label{e1}
\end{equation}

Also, according to \cite{34} the overall manufacturing cost of a typical DMFB is obtained using the following equation

\begin{equation}
Cost_{WR} = Cost_{PCB} + Cost_{SR}
\label{e2}
\end{equation}

According to above equation wire-routing cost of a typical DMFB is directly affected by the cost of PCB plus the cost of additional circuitry in the form of shift registers used for driving pins; additional circuitry is needed in case number of pins is higher than the number to be accommodated by the microcontroller so requiring shift registers. According to \cite{52}, using larger feature sizes tends to reduce the PCB cost. Equation \ref{e3} shows the elements effective on the PCB cost.

\begin{multline}
Cost_{PCB} = \\(numLayers, width_{PCB}, height_{PCB}, width_{WT})
\label{e3}
\end{multline}

\begin{figure}[b]
\centering
\includegraphics[scale=0.3]{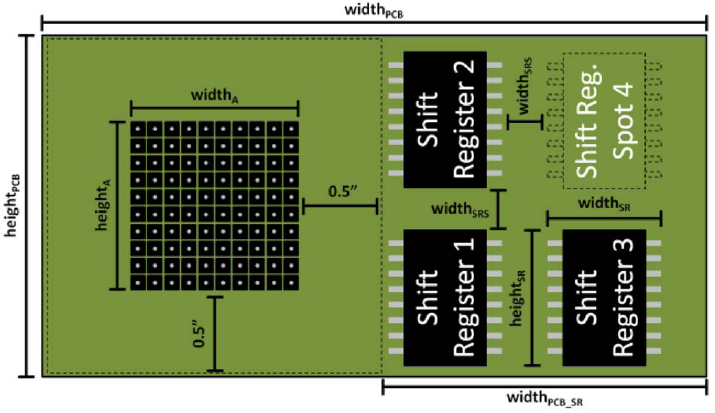}
\caption{DMFB layout for PCB size estimation \cite{34}}
\label{f13}
\end{figure}

Figure \ref{f13} shows the DMFB layout for PCB size estimation. As can be seen the array of electrodes is surrounded by a 0.5 inch perimeter of empty space; also the PCB width is extended to accommodate as many shift registers as necessary \cite{34}.

Table \ref{t2} defines parameters used in Figure \ref{f13}. As illustrated in the figure parameters width\textsubscript{PCB} and height\textsubscript{PCB} denote width and height of the PCB. Also, parameters width\textsubscript{A} and height\textsubscript{A} represent width and height of the array of electrodes. The parameters width\textsubscript{SR} and height\textsubscript{SR} correspond to width and height of the shift registers; additionally, parameter width\textsubscript{SRS} represents the spacing between shift registers.

Equations \ref{e4} and \ref{e5} calculate the PCB dimensions; also, the PCB space devoted to accommodation of shift registers is calculated through Equation \ref{e6}. According to \cite{34} shift registers are stacked vertically and in case necessary the width of PCB is increased to accommodate another column of shift registers.

\begin{equation}
\begin{IEEEeqnarraybox}[][c]{l?s}
\IEEEstrut
height_{PCB} = height_{Array} + 1 (inch)
\IEEEstrut
\end{IEEEeqnarraybox}
\label{e4}
\end{equation}

\begin{multline}
width_{PCB} = \\width_{Array} + width_{PCB-SR} + 1 (inch)
\label{e5}
\end{multline}

\begin{multline}
{\left \lceil \frac{\#ShiftRegs}{\left \lfloor height_{PCB} / (height_{SR} + width_{SRS}) \right \rfloor} \right \rceil} \\
\times (width_{SR} + width_{SRS})
\label{e6}
\end{multline}

\begin{table}[t]
\renewcommand{\arraystretch}{1.3}
\centering
\caption{PCB Parameters}
\label{t2}
\begin{tabular}{|l|c|}
\hline
\multicolumn{1}{|c|}{Feature} & Symbol \\ \hline
PCB Width & width\textsubscript{PCB} \\
PCB Height & height\textsubscript{PCB} \\
Array of Electrodes Width & width\textsubscript{Array} \\
Array of Electrodes Height & height\textsubscript{Array} \\
Shift Register Width & width\textsubscript{SR} \\
Shift Register Height & height\textsubscript{SR} \\
Shift Register Spacing Width & width\textsubscript{SRS} \\ \hline
\end{tabular}
\end{table}

Table \ref{t3} indicates PCB price estimates for varying number of layers and parameters. The figures have been obtained from Advanced Circuits’ online quote \cite{52}; the online quote system allows for specifying various parameters. For the sake of this study solely parameters of width and height of PCB, trace/size space, via size and number of layers were specified; leaving other parameters at their default values.

\begin{table}[b]
\renewcommand{\arraystretch}{1.3}
\setlength{\tabcolsep}{2.0pt}
\centering
\caption{PCB Cost Estimates for Varying Number of Layers and Parameters}
\label{t3}
\begin{tabular}{|c|c|c|c|c|c|ccccc|}
\hline
\multicolumn{2}{|c|}{\begin{tabular}[c]{@{}c@{}}Electrode\\ Pitch\end{tabular}} & \multicolumn{3}{c|}{\begin{tabular}[c]{@{}c@{}}Advanced Circuit\\ Metrics\end{tabular}} & & \multicolumn{5}{c|}{\begin{tabular}[c]{@{}c@{}}2\thinspace'' $\times$ 2\thinspace'' (@2,500 QTY)\\ with Varying Number of Layers\end{tabular}} \\ \cline{1-5} \cline{7-11} 
mm & in & \begin{tabular}[c]{@{}c@{}}Trace\\ Size/\\ Space\end{tabular} & \begin{tabular}[c]{@{}c@{}}Via\\ Size\end{tabular} & \begin{tabular}[c]{@{}c@{}}Via\\ Contact\\ Size\end{tabular} & \multirow{-2}{*}{\begin{tabular}[c]{@{}c@{}}\rotatebox[origin=c]{90}{O. Cap}\end{tabular}} & \multicolumn{1}{c|}{1} & \multicolumn{1}{c|}{2} & \multicolumn{1}{c|}{3} & \multicolumn{1}{c|}{4} & 5 \\ \hline
& & \cellcolor[HTML]{333333}{\color[HTML]{FFFFFF} 0.005} & \cellcolor[HTML]{333333}{\color[HTML]{FFFFFF} 0.010} & \cellcolor[HTML]{333333}{\color[HTML]{FFFFFF} 0.013} & \cellcolor[HTML]{333333}{\color[HTML]{FFFFFF} 2} & \cellcolor[HTML]{333333}{\color[HTML]{FFFFFF} \$1.20} & \cellcolor[HTML]{333333}{\color[HTML]{FFFFFF} \$1.20} & \cellcolor[HTML]{333333}{\color[HTML]{FFFFFF} \$1.73} & \cellcolor[HTML]{333333}{\color[HTML]{FFFFFF} \$1.81} & \cellcolor[HTML]{333333}{\color[HTML]{FFFFFF} \$2.09} \\
& & 0.004 & 0.008 & 0.011 & 3 & \$1.88 & \$1.88 & \$2.43 & \$2.52 & \$2.81 \\
& & 0.003 & 0.009 & 0.012 & 4 & N/A & N/A & N/A & N/A & N/A \\
\multirow{-4}{*}{1} & \multirow{-4}{*}{0.0394} & 0.002 & 0.008 & 0.011 & 5 & N/A & N/A & N/A & N/A & N/A \\ \hline
& & 0.008 & 0.028 & 0.038 & 2 & \$0.99 & \$0.99 & \$1.41 & \$1.46 & \$1.79 \\
& & 0.007 & 0.014 & 0.028 & 3 & \$0.99 & \$0.99 & \$1.41 & \$1.47 & \$1.79 \\
& & \cellcolor[HTML]{333333}{\color[HTML]{FFFFFF} 0.006} & \cellcolor[HTML]{333333}{\color[HTML]{FFFFFF} 0.012} & \cellcolor[HTML]{333333}{\color[HTML]{FFFFFF} 0.024} & \cellcolor[HTML]{333333}{\color[HTML]{FFFFFF} 4} & \cellcolor[HTML]{333333}{\color[HTML]{FFFFFF} \$0.99} & \cellcolor[HTML]{333333}{\color[HTML]{FFFFFF} \$0.99} & \cellcolor[HTML]{333333}{\color[HTML]{FFFFFF} \$1.46} & \cellcolor[HTML]{333333}{\color[HTML]{FFFFFF} \$1.53} & \cellcolor[HTML]{333333}{\color[HTML]{FFFFFF} \$1.80} \\
& & 0.005 & 0.013 & 0.023 & 5 & \$1.10 & \$1.10 & \$1.63 & \$1.71 & \$1.99 \\
\multirow{-5}{*}{2} & \multirow{-5}{*}{0.0787} & 0.004 & 0.008 & 0.026 & 6 & \$1.88 & \$1.88 & \$2.43 & \$2.52 & \$2.81 \\ \hline
& & 0.012 & 0.030 & 0.040 & 2 & \$0.99 & \$0.99 & \$1.41 & \$1.41 & \$1.79 \\
& & 0.010 & 0.020 & 0.030 & 3 & \$0.99 & \$0.99 & \$1.41 & \$1.41 & \$1.79 \\
& & 0.008 & 0.018 & 0.028 & 4 & \$0.99 & \$0.99 & \$1.41 & \$1.46 & \$1.79 \\
& & 0.006 & 0.024 & 0.034 & 5 & \$0.99 & \$0.99 & \$1.46 & \$1.53 & \$1.80 \\
& & \cellcolor[HTML]{333333}{\color[HTML]{FFFFFF} 0.006} & \cellcolor[HTML]{333333}{\color[HTML]{FFFFFF} 0.012} & \cellcolor[HTML]{333333}{\color[HTML]{FFFFFF} 0.022} & \cellcolor[HTML]{333333}{\color[HTML]{FFFFFF} 6} & \cellcolor[HTML]{333333}{\color[HTML]{FFFFFF} \$0.99} & \cellcolor[HTML]{333333}{\color[HTML]{FFFFFF} \$0.99} & \cellcolor[HTML]{333333}{\color[HTML]{FFFFFF} \$1.46} & \cellcolor[HTML]{333333}{\color[HTML]{FFFFFF} \$1.53} & \cellcolor[HTML]{333333}{\color[HTML]{FFFFFF} \$1.80} \\
& & 0.005 & 0.010 & 0.020 & 7 & \$1.20 & \$1.20 & \$1.73 & \$1.81 & \$2.09 \\
\multirow{-7}{*}{2.54} & \multirow{-7}{*}{0.1000} & 0.004 & 0.022 & 0.032 & 8 & \$1.38 & \$1.38 & \$1.93 & \$2.02 & \$2.31 \\ \hline
\end{tabular}
\end{table}

Looking at the figures provided in Table \ref{t3} PCB price estimates are provided for three electrode sizes of 1, 2 and 2.54 mm; within each category PCB price estimates are provided for different values of trace/size spacing, via size and number of layers. In every category of electrode sizes a single configuration with the most affordable pricing is highlighted in black.

Table \ref{t4} indicates the cost estimates of various sizes DMFB designs for the proposed GFPC and Enhanced FPPC architectures. The column ‘DMFB Details’ presents various information on the specifications of the design; the sub-column DMFB Name denotes name of DMFB design. The sub-columns \# Pins and \# SR represent number of control pins and shift registers required for driving pins of the DMFB design, respectively. Sub-column Adjusted PCB Dims represents the total size of PCB for accommodating the array of electrodes in addition to the number of shift registers (if any). The sub-column \# Layers shows the number of metal-layers used for wire-routing. The column Cost(\$) represents the cost estimation of the DMFB; the sub-columns Board and SR denote the cost estimation for the PCB (total size for accommodating the array of electrodes and any shift registers) and shift registers circuitry, respectively. The sub-column Total shows the sum of two sub-columns Board and SR, giving the total cost towards wire-routing of the DMFB.

\begin{table}[t]
\renewcommand{\arraystretch}{1.3}
\setlength{\tabcolsep}{3.0pt}
\centering
\caption{Cost estimates for various architecture sizes of Enhanced FPPC and the proposed Enhanced GFPC DMFB designs}
\label{t4}
\begin{tabular}{|c|c|c|c|c|c|c|c|c|}
\hline
\multicolumn{6}{|c|}{DMFB Details} & \multicolumn{3}{c|}{Cost (\$)} \\ \hline
\multirow{2}{*}{\begin{tabular}[c]{@{}c@{}}DMFB\\ Name\end{tabular}} & \multirow{2}{*}{\begin{tabular}[c]{@{}c@{}}\#\\ Pin\end{tabular}} & \multirow{2}{*}{\begin{tabular}[c]{@{}c@{}}\#\\ SR\end{tabular}} & \multicolumn{2}{c|}{\begin{tabular}[c]{@{}c@{}}Adjusted PCB\\ Dim.\end{tabular}} & \multirow{2}{*}{\begin{tabular}[c]{@{}c@{}}\#\\ Layers\end{tabular}} & \multirow{2}{*}{Board} & \multirow{2}{*}{SR} & \multirow{2}{*}{Total} \\ \cline{4-5}
& & & X (in) & Y (in) & & & & \\ \hline
EFPPC\_4 & 36 & 1 & 1.7638 & 1.6299 & 2 & \$1.01 & \$0.14 & \$1.15 \\
EGFPC\_4 & 36 & 1 & 1.8032 & 1.4331 & 2 & \$0.93 & \$0.14 & \$1.07 \\ \hline
EFPPC\_8 & 65 & 5 & 1.7638 & 2.1811 & 2 & \$1.16 & \$0.70 & \$1.86 \\
EGFPC\_6 & 52 & 3 & 2.0000 & 1.4330 & 3 & \$1.42 & \$0.42 & \$1.84 \\ \hline
\end{tabular}
\end{table}

Interpreting results obtained in Table \ref{t4} is seen that in case of 4 module variation of enhanced FPPC and the proposed design the enhanced FPPC design provides lower overall PCB costs given the lower layer-count; though, as the designs enlarge the proposed design would require smaller area and lower pin-count. Thus, in case the Enhanced FPPC design with 8 mixing modules and the proposed design with 6 modules the higher pin-count in the Enhanced FPPC design necessitates 5 shift registers for driving the controlling pins not to be accommodated by the microcontroller directly. Yet, in case of the proposed design the lower pin-count leads to lower number of shift registers thus incurring lower costs in terms of shift registers. The lower cost of shift register in case of the proposed design yields an overall cost lower than the enhanced FPPC design.

\section{PERFORMANCE SIMULATION RESULTS}

The performance simulation results are conducted using the UCR Static Synthesis Simulator \cite{20}, an open-source DMFB framework, provided by the researchers at the University of California Riverside.

Table \ref{t5} details performance simulation evaluation of conducting various bio-assays on the DMFBs. The bio-assays involve PCR (Polymerase Chain Reaction), In-vitro diagnostics (with variable number of samples and reagents), and Protein split; all of the bio-assays are available within the UCR SSS framework \cite{20}. The column ‘Name’ illustrates the name and type of the bio-assay. Column ‘Scheduling’ shows the scheduling time of the bio-assay; while column ‘Routing’ is devoted to the droplet routing times of the bio-assays. Column ‘Total’ is computed as the sum of the values of ‘Scheduling’ and ‘Routing’ columns which stands for the total time taken to perform the bio-assay.

\begin{table}[t]
\renewcommand{\arraystretch}{1.3}
\setlength{\tabcolsep}{6.0pt}
\centering
\caption{The performance simulation results (excluding wash droplets) of various bio-assays on the Enhanced FPPC (FP), the Proposed Design (PD)}
\label{t5}
\begin{tabular}{|l|c|c|c|c|c|c|}
\hline
\multicolumn{1}{|c|}{\multirow{2}{*}{Name}} & \multicolumn{2}{c|}{\begin{tabular}[c]{@{}c@{}}Scheduling\\ (s)\end{tabular}} & \multicolumn{2}{c|}{\begin{tabular}[c]{@{}c@{}}Routing\\ (s)\end{tabular}} & \multicolumn{2}{c|}{\begin{tabular}[c]{@{}c@{}}Total\\ (s)\end{tabular}} \\ \cline{2-7} 
\multicolumn{1}{|c|}{} & FP & PD & FP & PD & FP & PD \\ \hline
PCR & 11 & 11 & 2.2 & 1.7 & 13.2 & 12.7 \\
In-vitro\_1 & 14 & 14 & 3.4 & 1.9 & 17.4 & 15.9 \\
In-vitro\_2 & 16 & 18 & 5.0 & 2.8 & 21.0 & 20.8 \\
In-vitro\_3 & 16 & 18 & 7.6 & 4.5 & 23.6 & 22.5 \\
In-vitro\_4 & 18 & 19 & 10.1 & 6.4 & 28.1 & 25.4 \\
In-vitro\_5 & 21 & 25 & 13.2 & 9.1 & 34.2 & 34.1 \\
Protein\_split\_1 & 52 & 52 & 2.7 & 1.6 & 54.7 & 53.6 \\
Protein\_split\_2 & 62 & 62 & 8.4 & 7.4 & 70.4 & 69.4 \\
Protein\_split\_3 & 83 & 83 & 18.2 & 16.0 & 101.2 & 99.0 \\ \hline
\multicolumn{1}{|c|}{Average} & \multicolumn{2}{c|}{- 3\%} & \multicolumn{2}{c|}{+ 27\%} & \multicolumn{2}{c|}{+ 3\%} \\ \hline
\end{tabular}
\end{table}

\begin{table}[h]
\renewcommand{\arraystretch}{1.3}
\setlength{\tabcolsep}{6.0pt}
\centering
\caption{The performance simulation results (including wash droplets) of various bio-assays on the Enhanced FPPC (FP), the Proposed Design (PD)}
\label{t6}
\begin{tabular}{|l|c|c|c|c|c|c|}
\hline
\multicolumn{1}{|c|}{\multirow{2}{*}{Name}} & \multicolumn{2}{c|}{\begin{tabular}[c]{@{}c@{}}Scheduling\\ (s)\end{tabular}} & \multicolumn{2}{c|}{\begin{tabular}[c]{@{}c@{}}Routing\\ (s)\end{tabular}} & \multicolumn{2}{c|}{\begin{tabular}[c]{@{}c@{}}Total\\ (s)\end{tabular}} \\ \cline{2-7} 
\multicolumn{1}{|c|}{} & FP & PD & FP & PD & FP & PD \\ \hline
PCR & 11 & 11 & 2.2 & 1.7 & 13.2 & 12.7 \\
In-vitro\_1 & 14 & 14 & 3.4 & 1.9 & 17.4 & 15.9 \\
In-vitro\_2 & 16 & 18 & 5.0 & 2.8 & 21.0 & 20.8 \\
In-vitro\_3 & 16 & 18 & 7.6 & 4.5 & 23.6 & 22.5 \\
In-vitro\_4 & 18 & 19 & 10.1 & 6.4 & 28.1 & 25.4 \\
In-vitro\_5 & 21 & 25 & 13.2 & 9.1 & 34.2 & 34.1 \\
Protein\_split\_1 & 52 & 52 & 2.7 & 1.6 & 54.7 & 53.6 \\
Protein\_split\_2 & 62 & 62 & 8.4 & 7.4 & 70.4 & 69.4 \\
Protein\_split\_3 & 83 & 83 & 18.2 & 16.0 & 101.2 & 99.0 \\ \hline
\multicolumn{1}{|c|}{Average} & \multicolumn{2}{c|}{- 3\%} & \multicolumn{2}{c|}{+ 27\%} & \multicolumn{2}{c|}{+ 3\%} \\ \hline
\end{tabular}
\end{table}

Looking at the figures provided in Table \ref{t5} it is seen that the scheduling times of the enhanced FPPC is superior in case of In-vitro\_2 to In-vitro\_5 bioassays; thereby outperforming the proposed design by 3 percent. Next, considering droplet routing times it is noted that the proposed design performs remarkably lower droplet routing times compared with the enhanced FPPC design; this is due to architecture of the proposed design and also smaller size of the architecture compared with the enhanced FPPC design.

\section{Conclusion}

In this paper an enhanced general-purpose field-programmable pin-constrained DMFB design was proposed. The enhanced design was aimed at reducing the manufacturing costs while retaining and improving the overall performance. According to cost analysis and performance simulation results the proposed design demanded lower PCB costs and also at the same time provided faster total bioassay execution times in average.

\end{document}